\documentstyle[version2,preprint,aps]{revtex}
\begin{document}
\draft

\begin{titlepage}

\begin{title}
Correlations in a Many-Body Calculation of $^{\bf 11}$Li
\end{title}

\author{ C. R. Chinn}
\begin{instit}
Department of Physics and Astronomy \\
Vanderbilt University,
Nashville, TN  37235 \\
\end{instit}

\author{J.~Decharg\'e and J.-F.~Berger}
\begin{instit}
{\it Service de Physique et Techniques Nucl\'{e}aires \\
 Centre d'Etudes de Bruy\`{e}res-le-Ch\^{a}tel}\\
{\it B.P. No. 12, 91680 Bruy\`{e}res-le-Ch\^{a}tel, FRANCE} \\
\end{instit}

\begin{center}
{\bf{Abstract}}
\end{center}

\begin{abstract}
A many-body calculation of $^{11}$Li is presented where the only
input is the well-tested, finite-range {\it D1S} effective interaction
of {\it Gogny}.  Pairing correlations are included in a
constrained Hartree-Fock-Bogolyubov calculation, while long-range collective
correlations are introduced using a GCM derived calculation.
Correlations are found to play an important role in describing
$^{11}$Li.  A substantive underlying $^9$Li core of $^{11}$Li is
found, which has a different density
profile than a free $^9$Li nucleus.  This may have significant
implications in the use of a three-body framework in studies of
$^{11}$Li.
\end{abstract}

\end{titlepage}

\pagebreak


\newcommand\lsim{\thinspace{\hbox to 8pt{\raise
-5pt\hbox{$\sim$}\hss{$<$}}}\thinspace}
\newcommand\rsim{\thinspace{\hbox to 8pt{\raise
-5pt\hbox{$\sim$}\hss{$>$}}}\thinspace}

\section{Introduction}

With the recent advent of secondary beam facilities there has been
a large research interest in nuclei near the drip lines.  These
exotic nuclei offer opportunities to study many-body effects under
unusual conditions.  An example of such can be related to the
nucleon-nucleon [N-N] interaction.  While the free N-N interaction
appears to be well understood, the role of the N-N interaction in
microscopic nuclear structure is far from being clear.  (One
example where one would think that the role of the N-N interaction
should be clear, but is not is in the $^3$H problem.)  Because
nuclei near the drip line have weak binding energies and hence
large density distributions, the N-N interaction can now be
studied in regions of low nuclear density.

The free neutron-neutron [n-n] interaction is attractive, but the
di-neutron system is unbound.  Migdal and Watson postulated the
possibility that a di-neutron may become bound if placed within
the field of a nucleus.  $^{11}$Li among others appears to be such
a system, since both $^{11}$Li and $^9$Li are bound while $^{10}$Li
is not.  Hence, with this in mind, recent interest in $^{11}$Li
has been strongly addressed within a three-body framework
\cite{bang1,bang,suzuki,jensen,hansen,Hayes}, where
$^{11}$Li is represented as two neutrons surrounding an
inert $^9$Li core.  For a recent review of such work, please see
Ref.~\cite{bang1}.

The assumption that $^{11}$Li can be realistically represented as
a three-body system needs to be investigated.
While present experimental evidence indicates that this assumption
may be valid \cite{tani,exp}, one would like to
test this hypothesis within a many-body theoretical investigation.
Difficulties that may arise with the three-body framework include:
1.) treatment of the Pauli exclusion principle can only be
performed approximately, and 2.)  correlations with the $^9$Li core
are ignored.

The most straightforward way to study $^{11}$Li in an $A$-body
framework is to use a single particle model, but this was
shown for this nucleus \cite{bertscha} to give an inadequate description.
Therefore, for a proper $A$-body calculation a correlated
description beyond a single particle model is required.  By
renormalizing the mean field potential several groups
\cite{bertscha,sagawa,zhu} found that the experimental two neutron
separation energy and the $rms$ radius of $^{11}$Li could be
reproduced, thus indicating the possible existence of
correlations among the outer single particle states.  An attempt
to model correlations in a shell model description \cite{Arima}
resulted in small effects, but as will be shown later in this
paper the adopted shell model space used in this study is
clearly too small.

In this paper we introduce long-range correlations by using a Generator
Coordinate Method [GCM] type formalism. The
nuclear ground state [GS] is
represented as a superposition of HFB nuclear states, which are
obtained by constraining on different values of the
$\langle r^2\rangle$ collective variable. The reason for choosing this
particular GCM variable can be understood in the following way:
correlations in the $^{11}$Li GS are expected to occur because the loosely
bound outer neutrons can occupy a large number of nearly degenerate
Rydberg-type orbits having a broad range of radial extensions. One
expects that a constraint imposing different values of the total $rms$
radius will act essentially on the outer neutrons radial distributions
and therefore will be able to generate the kind of configurations
present in the nuclear GS. Therefore, as is usual in GCM
calculations, initially, a series of constrained HFB microscopic mean
field calculations is performed. In a second step the coefficients of
the GCM configuration mixing are computed by solving the coupled
equations resulting from the application of a variation principle to
the total nucleus binding energy.  In the present work, this is
accomplished in an approximate fashion by reducing the usual
Hill-Wheeler equations to a collective Schr\"odinger equation
of the Bohr type. Note that this calculation includes pairing
correlations as well as long-range collective degrees
of freedom as derived from the effective N-N interaction. Also,
the Pauli principle is strictly obeyed through the use of fully
antisymmetrized single particle wave functions.

 The only input into this consistent $A$-body calculation is the
well-tested {\it D1S}  {\it Gogny} force. This interaction is a
density-dependent phenomenological parametrization of the N-N
interaction inside the nuclear medium which includes a spin-orbit term,
and is finite-ranged. The parameters of the interaction have been fixed
by matching the bulk properties of nuclear matter and of a few finite
nuclei, including pairing correlation strengths. This {\it D1S} force has
been tested in a variety of applications with excellent results
\cite{gogny,HFB}. It must be noted that this force gives a very
good description not only of medium and heavy nuclei, but also of very
light nuclei. For instance it describes correctly the binding energy
and radius of the alpha particle. In addition the finite range,
density-independent part of the force has been set-up in order to
roughly simulate a free N-N interaction in the sense that it gives the
correct N-N scattering lengths. These properties are important in the
present context which deals with a three-proton system where almost
free neutron-neutron interactions are expected to play a crucial
role.

With this collective model a qualitative study of correlations
in $^{11}$Li is presented.  In particular the role of the $^9$Li
core is explored, especially its relation to a free $^9$Li
nucleus.  The validity of the three-body hypothesis is
investigated using a collective $A$-body model.

In Section~II the mean field constrained and unconstrained HFB
calculation is described and the results presented.
Long-range correlations using a
simplification of the GCM is described in Section~III along with
the results and analysis, followed by a conclusion.

\section{Constrained HFB Calculation}

     The constrained HFB $A$-body calculation is performed using a nineteen
shell axially-symmetric harmonic oscillator [HO] basis.  In performing
tests of the convergence of the basis, it was found that a
nineteen shell basis is required due to the large extension of the
neutron matter distribution from the center of the nucleus in
coordinate space, and to the inadequate asymptotic behavior of HO
states in $r$ space.  The use of a multi-oscillator basis,
{\it i.e.} of a basis composed of several sets of concentric HO states
associated with different lengths is also used in the HFB calculation. This
kind of basis allows one to extend to larger distances the radial description
of nucleon orbits. However, in this case only spherically symmetric nuclear
distributions could be described.
Time-reversal symmetry is assumed, so the
protons are described by blocking with equal weights the
two $j_z^\pi = \pm 3/2 ^-$ axial quasi-particle orbitals, thus
matching the known ground state spin of both $^{11}$Li and
$^{9}$Li.

For a study of a small nucleus such as $^{11}$Li, it would be
expected that a mean field description would not be the most
appropriate choice.  In this case we wish to address certain
many-body questions and to test assumptions about the many-body
nature of the problem.  With this in mind as a first step, a mean
field calculation using HFB should be able to address some of
these qualitative concerns.

As explained in the Introduction, the $^{11}$Li nucleus is expected
to be very soft against changes
in the density distribution $rms$.  For this reason a constrained
HFB calculation was performed, where the constraint variable
used corresponds to the mean value of $\langle r^2\rangle$:
\begin{eqnarray}
q & = & \int dr\, r^2\, \rho(r) ~,~{\rm where} ~
    \rho(r) = \int d\Omega \, \rho({\vec r}) ~,~ \\
\langle r_{rms}\rangle & = & \sqrt{q} \nonumber
\end{eqnarray}

When the constraint is switched off, one gets the mean-field
representation of the $^{11}$Li GS. The total $rms$ matter radius obtained
in this way is about 2.80~fm, matching the results of several
other groups \cite{Arima,bertscha}.  The separate proton and neutron
$rms$ radii are found to be 2.30 and 2.97~fm, respectively.  For a
similar HFB calculation of the $^{9}$Li GS the $rms$ radii are found to
be 2.47, 2.24, and 2.58~fm for the total, proton and neutron
distributions, respectively.  The $rms$ radii for the protons in
$^{9}$Li and $^{11}$Li differ only by a small amount in these
mean-field calculations, resulting from the effects of the force
between the neutrons and the protons.  However, when correlations are
introduced in $^{11}$Li,  if there is an uncorrelated $^{9}$Li core in
$^{11}$Li , it may well differ from the free $^{9}$Li GS.  It remains
to be seen though how large this difference may be.

In Fig.~1 the results of the constrained $^{11}$Li calculation are
shown as a function of the constraint variable, $q$.  A similar
spherical $^9$Li calculation is also shown.  As $q$ is increased,
the $^{11}$Li curve is much softer than $^9$Li in the sense that
the slope is much less steep.  One then expects $^{11}$Li to
include significantly more configuration mixing than $^9$Li.  Such a
mixing has been included and will
be discussed in detail later.

The $rms$ radii for the neutron and proton distributions are shown
in Fig.~2a for the constrained HFB results as a function of
$\langle r_{rms}\rangle_{total}=\sqrt{q}$.  As expected (see the
Introduction) the proton $rms$
is  unaffected by the constraint, while the neutron
$\langle r_{rms}\rangle_n$ varies linearly.  In other words, the curve
confirms that the potential between
the protons and outer neutrons  is not strong
enough to prevent the two distributions from decoupling.

The independent nature of the proton and neutron sectors is
also evident in the pairing energy shown in Fig.~2b. The
protons consistently have zero pairing as a function of $q$.  The
neutrons have strong pairing for  $q > 3$~fm indicating the onset of
a significantly high neutron level density at the Fermi surface.

Since the protons do not exhibit pairing there must be a sizeable
gap at the Fermi surface.  This is shown in Fig.~3a, where the
protons occupying the $1s1/2$ and $1p3/2$ states.  There is a gap
of about $6$~MeV between the $1p3/2$ level and the higher single
particle levels and this gap remains for all $q$ considered.  The
relative energies of the occupied levels do not change
significantly as a function of $q$, confirming the negligible
influence of the constraint on the
proton mean-field.  Note that, since the $1p3/2$ level does not shift a
great deal, the blocking approximation used here to account for the odd
number of protons should be reasonable.

In Fig.~3b the corresponding neutron single particle levels are
shown.  The lowest six neutrons are almost completely contained in
the $1s1/2$ and $1p3/2$ orbitals.  Due to pairing correlations the
last two neutrons are dispersed throughout the higher levels.
There is almost no gap between the $1p1/2$ level and the higher
levels, and for large $q$ these levels cross.  The levels higher than
$2d3/2$ are not shown.  The level density at the Fermi surface is
very high, indicating that configuration mixing, {\it i.e.} the
existence of correlations is widespread.

This result is an indication that the model space required to describe
the $^{11}$Li GS in an extended shell model calculation is much
larger than generally assumed~\cite{Arima}. According to the present
calculation levels above the $f$~shell surely contribute to the GS
description.

In Fig.~3b it is clear that there is a large gap between the
six inner neutrons, which represent a $^9$Li core, and the outer
two valence neutrons.  The calculated occupation probabilities for
the core neutrons are almost always between $1.00$ and $0.99$ with
a minimum of $\rsim 0.987$.  This is true for all $q$ considered.
Because of these features, a $^9$Li core wave function can be
projected out by taking the constrained HFB $^{11}$Li solutions
and explicitly setting the first six neutron levels to have one
occupation probability and the other neutron levels to be empty.
As evidenced by the fact that the core
neutron probabilities are not exactly one, this is an
approximate procedure, but clearly, because of the large energy
gap, this should be a very reasonable and accurate representation.

In Fig.~4 the neutron density profiles are shown for the
unconstrained HFB calculations of  $^9$Li and $^{11}$Li, and for
the $^9$Li
core projected from  $^{11}$Li as explained above.  At the center of
the nucleus the free $^9$Li neutron density is sizeably larger than the
$^9$Li core neutron density, and accordingly, the $^9$Li core extends
somewhat farther.  More precisely, the central neutron density of
$^{9}$Li (short-dashed curve) is about 15\% larger than that of
the $^{9}$Li core (long dashed curve).  A similar difference
was found for the protons.  This is more easily seen when the
tail is expanded in a logarithmic plot. The the $^{11}$Li core
neutrons density doesn't fall off as rapidly beyond $r = 7$~fm.
Already one can see evidence for the beginning of a halo-like
structure.

\section{Calculation Including Configuration Mixing}

{}From the evidence shown in the previous two sections it is clear
that a pure single particle model is inadequate to describe
$^{11}$Li.  To provide a more sophisticated representation, a
correlated ground state wave function is constructed as a
superposition of the HFB nuclear states in the following GCM
form:
\begin{equation}
|\chi_o\rangle = \int dq \, f_o(q) \, |\phi_q\rangle ~,  \label{eq:3}
\end{equation}
where $|\phi_q\rangle$ is a product of HFB quasiparticle states
for deformation $q$, $f_o(q)$ is a weight function, and $q$ is
the constraint variable.  In the GCM formalism a hamiltonian kernel
is constructed with the GCM wave function.  By applying the
variational principle, an equation is derived from which the weight
functions can be calculated.  A Gaussian overlap approximation
[GOA], where the overlap between any two HFB states is approximated
by a Gaussian in the collective variable \cite{ring}, is applied
to simplify these equations, deriving a Bohr Hamiltonian
expression.  The solution of the Bohr Hamiltonian equation gives
the weight function, $f_o(q)$.  These techniques have been
thoroughly tested in many instances \cite{isomer}.
It may be that the GOA is not
as accurate for smaller nuclei as in previous experience,
but for the investigative study being performed here, this should
be more than adequate.

The resulting ground state collective wave function,
$f_o(q)$, has a corresponding total binding energy.  To properly
calculate the two neutron separation energy it is necessary to
perform a similar collective calculation for the free $^9$Li case
and then to take the difference in the total collective
binding energies.  As
shown in Ref.~\cite{bertscha} the energies calculated from the
single particle HFB model are inadequate and it is necessary to
include more sophisticated degrees of freedom.

The ground state correlated density is calculated from (\ref{eq:3})
in the following fashion:
\begin{equation}
\rho_{GS}(r) = \int dq' \, \int dq \, f_o(q') \, \langle\phi_{q'} | \,
{\hat \rho}_o(r) \, |\phi_q\rangle \, f_o(q) , \label{eq:4}
\end{equation}
where ${\hat \rho}_o(r)$ is the radial density operator
obtained after angle averaging.

In Fig.~5 the proton density profile is shown from the
unconstrained HFB calculation of $^9$Li and $^{11}$Li along with
the GCM-type collective result for $^{11}$Li.  As is apparent here
the $^9$Li and $^{11}$Li proton profiles are not equivalent,
implying differences between the free $^9$Li nucleus and
the $^9$Li core of $^{11}$Li.  The correlated $^{11}$Li
proton density is very similar to the uncorrelated $^{11}$Li
result, which should not be surprising in light of
the unchanging single particle spectrum in Fig.~3a.
For the protons the correlations have little effect upon the
density distributions both near the central part and along the
tail.  It appears clear that there are few correlations in
the $^{11}$Li proton sector.

The various neutron distributions are shown in Fig.~6.
As in the unconstrained HFB case an approximate representation
of the $^9$Li core can be projected
out from the $^{11}$Li calculation including configuration mixing.
This is performed by setting the neutron occupation propabilities
in the constrained HFB solutions to be either one or zero and then
using these density matrices with the weight function, $f_o(q)$,
obtained in the collective $^{11}$Li calculation.
As for the proton case the influence of correlations on the $^9$Li
core appears quite small. The collective long-range correlations
represented by the GCM therefore have little effect on the $^9$Li
core as a whole.

This influence is much bigger on the full $^{11}$Li neutron
distribution, that is when one includes the two extra neutrons.
The surface of the neutron distribution of
$^{11}$Li is at about $\sim 2.5$~fm, which is much further
extended than the surface of the proton distribution
at $\sim 1.8$~fm. Correlations slightly reduce this difference.
Substracting the  $^9$Li core neutron distribution from the
$^{11}$Li neutrons gives the structure of the two valence
neutrons.  The two neutrons have zero density at the origin due to
the Pauli blocking from the core and extend in a halo-like
structure.  At the surface the collective valence structure causes
the collective calculation to have a different neutron density
profile than the unconstrained HFB neutrons.

This is most easily seen on the logarithmic plot in the lower
part of Fig.~6 :  the HFB and GCM GS $^{11}$Li neutron densities
strongly differ beyond $q=6$~fm. At this point
it must be emphasized that an unphysical ledge
appears in the GCM densities beyond $q=7-8$~fm. This ledge is small
in amplitude, but it affects greatly the calculated $rms$ radius
since it extends very far out. The
origin of this ledge can be traced back to the structure of the
HFB constrained solutions included in the configuration mixing.
In Fig.~7 the neutron densities obtained in constrained HFB
calculations using the nineteen shell basis at $q= 3.4$~fm
and $q=4$~fm are
shown.  Also shown are the same results using a multi-oscillator
spherical HFB basis.  This particular multi-oscillator basis uses
three concentric sets of eight shell bases with three different
oscillator lengths.
This basis corresponds to a very large single oscillator basis
($\approx 60$~shells).  Clearly, a ledge appears in the HFB densities
which depends on the basis. With the large multi-oscillator basis the
ledge appears two orders of magnitude smaller and at larger $r$.  This
ledge is a result of the constraint on $\langle r^2\rangle$, that
tends to push up
the density at large $r$, but then is restricted by the local
nature of the harmonic oscillator bases. One in
fact observes the parabolic fall of the densities in the 19 shell HO
basis, characteristic of the HO asymptotic behavior at large $r$.
A similar parabolic fall of the multi-oscillator densities
is also observed at very large $r$ (30 fm). Both the property of
$r^2$ to be unbounded and the use of a restricted HO space may be
responsible for this phenomenon

Since there clearly exists a great deal of configuration mixing
in $^{11}$Li due to the radial extension of outer neutrons, one
would also like to be able to
extract a reasonable asymptotic tail for the GS wave function. This
is necessary to get a reasonable estimate of the GS
neutron $rms$ radius. The single particle wave functions should
asymptotically be proportional to:
\begin{equation}
 \sim~ \frac{e^{-\kappa r}}{r} ~,~~~
     \kappa = \sqrt{\frac{-2\mu E}{\hbar^2}}  .
\end{equation}
where $E$ is the single particle energy. For large but finite $r$ a
polynomial in $1/r$ should be included. Therefore, one expects
the HFB neutron density to behave for large $r$,  $r>r_o$ say, as:
\begin{equation}
 \langle q|\rho(r)|q\rangle = \rho_{o}(q) \frac{e^{-2\alpha_q r}}
   {r^2} \left[ 1 + \frac{b_q}{r} + \frac{c_q}{r^2}\right]^2 ,
  \label{eq:6}
\end{equation}
This form has been used to extrapolate the densities obtained
in the multi-oscillator basis for values of $r$ beyond the values
where the unphysical ledge begins.
The parameters $\alpha_q$, $b_q$ and $c_q$ were
obtained by using a $\chi^2$ fitting routine.  $\rho_{o}(q)$ was
chosen to match the small $r$ density profile at $r_o$, where the
tail was attached to the single oscillator calculation, typically
at about $7$~fm.  The off-diagonal terms in the correlated
densities (3) could then be computed from:
\begin{equation}
 \langle q'|\rho(r)|q\rangle \propto
\frac{e^{-\left(\alpha_{q'}+\alpha_{q}\right) r}}{r^2}
  \left[ 1 + \frac{b_{q'}}{r} + \frac{c_{q'}}{r^2}\right]
  \left[ 1 + \frac{b_q}{r} + \frac{c_q}{r^2}\right] .
\end{equation}
There is a great deal of freedom in the choice of the tail
parameters and hence we were able to obtain various different
parameter sets, depending on how we chose to fix things.  The
$\alpha_q$ parameters were set to be between
$0.10 \rightarrow 0.15$, which corresponds to energies of about
$-0.21 \rightarrow -0.47$~MeV.

With these fitted tails Fig.~8 is obtained.  Different sets of
fitted tail parameters are labeled by a, b, c, d, e. They give
slightly but not very different values for
the extrapolated density above  $10^{-6}$.  Note that here the
tails extend further out than for the unconstrained HFB case.

Let us now turn to result concerning  $rms$ radii.  In Table~I
the $rms$ radii for the various densities of figs. 4, 5 and 6
are shown.  The correlations change the calculated $rms$ radii
for the $^{9}$Li core by only about 2\%, which
is consistent with the previous observation that correlations have a
negligible effect upon the core. In this change the tail correction
made above plays almost no role.  As to the
differences in the $rms$ radii between the $^{9}$Li nucleus
and the values calculated for the $^{9}$Li core, they are about 5\%
for both the protons and the neutrons. This effect upon the
$rms$ radius appears to be small, but the comparison of the density
distributions made in the previous section yields a more pronounced
difference.  One may say that the  $^{9}$Li system is slightly
inflated when two extra neutron are added. This comparison assumes
that the $^{9}$Li nucleus ground state wave function is
not significantly influenced by correlations, which certainly is
reasonable in view of the insensitivity of correlations
on the  $^{9}$Li core. Again, the comparisons made for the
nine-nucleon
systems don't depend on the tail correction introduced previously.
One may conclude that in this many-body calculation of
$^{11}$Li there appears to exist a substantive $^{9}$Li core
which notably differs from a free $^{9}$Li nucleus.  One
expects that these differences may have a nonnegligible
effect upon the mean field felt by the two outer neutrons.

Going to the full $^{11}$Li $rms$ radii, the value listed in
Table~I with correlations included, but without tail correction
(3.42 fm), clearly is overestimated.  In
Table~II the $rms$ radii obtained without and with the various
parametrized tail corrections  are listed.
The largest $\langle r_{rms}\rangle_{tot}$ using a fitted
tail is case~$a$ with $2.88$~fm.  In the calculations performed
here, the increase in $rms$ radii due to correlations is between
$\sim 0.04 \rightarrow 0.08$~fm with realistic tails. This result is
somewhat disappointing in view of the the amount of complexity put
into the GS wave-functions. The total and neutron $rms$ radii,
although not so far from consistent with the experimental error
bars, appear $0.2$~fm
smaller than the nominal experimentally extracted values.

At this point, one may note that the present radius results probably
represent lowest values. In fact, when using the multi-oscillator basis
the resulting potential energy surface appears somewhat flatter than the
nineteen shell basis shown in Fig.~1. This indicates that that a
fully realistic calculation where density tails would be correctly
described, would also yield a much softer collective potential and
therefore a stronger configuration mixing. Hence effects of
correlations larger than those derived here, especially with respect to
the $rms$ radius predictions, would certainly be found.

Finally let us mention that, since the present $^{11}$Li
calculations were performed assuming axial symmetry,  we were
able to calculate the electric quadrupole moments, both in the
HFB and correlated calculations.  It should be realized that since
the proton sector was calculated with a blocking approximation,
the ground state spin of the nucleus has been artificially set.
In Table~III the calculated axial quadrupole moments are displayed
for $^{11}$Li and $^9$Li.  Since the proton sector is not greatly
affected by the correlations, the HFB and correlated GCM
calculations give approximately the same charge quadrupole moment.
Both of these results give remarkable agreement with the
experimental result.  The free $^9$Li HFB result is also shown,
where here the agreement is not nearly as close.

\section{Conclusions}

A microscopic many-body calculation of $^{11}$Li is presented with
correlations.  Pairing correlations are included in a constrained
HFB calculation, while important long-range correlations are
incorporated using a GCM type formalism.  The sole input into the
calculation is the well-tested, finite-range {\it D1S} effective
interaction.

It was found that long-range correlations play an important role
in the description of $^{11}$Li.  A distinct $^9$Li core was found
which remained uncoupled to both pairing and long-range
correlations.  The correlations were restricted almost exclusively
to the sector occupied by the two valence neutrons.
This substantive core appears to support the use of
a three-body framework to study this nucleus, although several
factors must be considered.  The Pauli principle acting between
the $^9$Li core neutrons and the two valence neutrons must be
taken into account as seen in Fig.~6.  Also, the $^9$Li core
found here appears to have a significantly different
density profile than a free $^9$Li nucleus.  It may be necessary
for three-body calculations to take into account the different
density shape of the $^9$Li core if more accurate calculations
of this type are desired than those presently in existence.

Difficulties in this many-body treatment of $^{11}$Li were
encountered due to the use of naturally localized harmonic
oscillator bases.  Even with the use of an extended
multi-oscillators dependences upon the basis were found.
For a more complete and accurate calculation it would be necessary
to use a better basis representation, which can accomodate a
nucleus with a large extended tail.  For example a collocation Basis
spline basis in a very large box would probably work well.

\vfill
\vspace{10mm}
\noindent
{\bf Acknowledgements:  }Appreciation is extended to J.~Luis~Egido
for bringing our attention to this problem and to M.~Girod,
V.~Khodel and R.~M.~Thaler for valuable discussions.  One of the
authors (CRC) would like to acknowledge the gracious hospitality
of the Service de Physique et Techniques Nucl\'eaires at
Bruy\`eres-le-Ch\^atel where this research was performed.


\pagebreak



\vfill
\pagebreak

\figure {The Total Hartree-Fock-Bogolyubov energy as a function
         of the constraint variable, $\langle r^2\rangle$ for
         $^{11}$Li and $^{9}$Li is shown.  The $^{9}$Li resulted is
         taken from a spherically symmetric HFB calculation. }

\figure { In the upper panel the separate $rms$ radii for the proton
          and neutron distributions is plotted for the constrained
          HFB calculation as a function of the square root of the
          constraint variable, the $rms$ radii of the total matter
          distribution.  The lower panel plots in a similar fashion
          the BCS pairing energies for the neutron and proton
          sectors.  }

\figure { The energies for some of the lowest single particle states are
          plotted for the protons and neutrons in the upper and
          lower panels, respectively, for the constrained HFB
          calculation as a function of $\langle r^2\rangle$.  }

\figure { The neutron radial densities for the unconstrained HFB
          calculation are shown as a function of $r$.
          The solid and short-dashed curves correspond to the
          calculated neutron densities of $^{11}$Li and $^9$Li,
          respectively.  The long-dashed curve represents the
          neutron density for the $^9$Li core projected from
          $^{11}$Li.  The lower panel is a logarithmic plot of the
          upper panel with the x-axis extended.  }

\figure { The proton radial density profiles are shown as a function of
          $r$.  The solid and short-dashed curves correspond to
          the unconstrained HFB calculation for $^{11}$Li and
          $^9$Li, respectively.  The long-dashed curve is the
          result from the correlated GCM-type calculation.  The
          lower panel is a logarithmic plot of the
          upper panel with the x-axis extended.  }

\figure { The neutron radial density profiles are shown as a
          function of $r$.  The solid and short-dashed curves
          correspond to the unconstrained HFB calculation of
          $^{11}$Li and the projected $^9$Li core of $^{11}$Li,
          respectively.  The long-dashed and dot-dashed curves
          are the result from the correlated GCM-type calculation
          for $^{11}$Li and the projected $^9$Li core of
          $^{11}$Li, respectively.  The shaded, fat long-dashed curve
          represents the difference between the long-dashed and the
          dot-dashed curves, thus corresponding to the two outer
          valence neutrons.  The lower panel is a logarithmic plot
          of the upper panel with the x-axis extended.  }

\figure {The neutron radial densities are shown for the
         constrained HFB calculation in a logarithmic plot as a
         function of $r$.  The solid and long-dashed curves
         correspond to calculations using a 19 shell harmonic
         oscillator basis, where $q$ is constrained to $3.4^2$
         and $4.0^2$~fm$^2$, respectively.  The short-dashed and
         dot-dashed curves use three concentric eight shell harmonic
         oscillators as a basis, where $q$ is constrained to
         $3.4^2$ and $4.0^2$~fm$^2$, respectively.   }

\figure { The collective neutron radial densities from the
          GCM-type calculation are shown, where the artificial
          tails of various listed categories were used for the
          densities calculated in the constrained HFB calculations.  }

\begin{tabular}{|c|ccc|}
\hline
  Type of Calculation & $\langle r_{\rm total} \rangle$ [fm]~~ &
  $\langle r_{\rm proton} \rangle$ [fm]~~ &
  $\langle r_{\rm neutron} \rangle$ [fm] \\[2mm] \hline
  $^{11}$Li nucleus, unconstrained HFB & 2.80 & 2.30 & 2.97 \\
  $^{11}$Li nucleus with correlations & 3.42 & 2.31 & 3.75   \\
  $^{9}$Li core projected from ~~~~~~~~~~~& & & \\ [-5mm]
  $^{11}$Li in unconstrained HFB  & 2.55 & 2.30 & 2.67 \\
  $^{9}$Li core projected from ~~~~~~~~~~~& & & \\ [-5mm]
  $^{11}$Li with correlations  & 2.59 & 2.31 & 2.72 \\
  $^{9}$Li nucleus, unconstrained HFB & 2.47 & 2.24 & 2.58 \\
  $^{11}$Li , experiment \cite{tani} & 3.12(.16) & 2.88(.11) & 3.21(.17) \\
  $^{9}$Li , experiment & 2.32(.02) & 2.18(.02) & 2.39(.02) \\ [2mm] \hline
\end{tabular}
\noindent

\begin{itemize}
\item[Table I:]  The calculated $rms$ radii for the neutron, proton and total
           matter distributions of $^{11}$Li, $^{9}$Li and the $^{9}$Li
           core projected from $^{11}$Li calculation are listed.  The results
           from the pure mean field HFB and the correlated calculations
           are used.
\end{itemize}

\begin{tabular}{|c|ccc|}
\hline
  With fitted tail & ~~~$\langle r_{\rm tot} \rangle$~~~  &
    ~~~$\langle r_{\rm prot} \rangle$~~~  &
    ~~~$\langle r_{\rm neut} \rangle$~~~ \\[2mm] \hline
  $^{11}$Li nucleus, HFB & 2.80 & 2.30 & 2.96 \\
  $^{11}$Li w/correlations & 3.42 & 2.31 & 3.75   \\
  $^{11}$Li Case a & 2.88 & 2.31 & 3.07 \\
  $^{11}$Li Case b & 2.84 & 2.31 & 3.02 \\
  $^{11}$Li Case c & 2.85 & 2.31 & 3.03 \\
  $^{11}$Li Case d & 2.86 & 2.31 & 3.04 \\
  $^{11}$Li Case e & 2.87 & 2.31 & 3.05 \\ [2mm] \hline
\end{tabular}
\noindent

\begin{itemize}
\item[Table II:]  The calculated $rms$ radii in fm for the neutron,
                  proton and total matter distributions of $^{11}$Li
                  including correlations are tabulated using various
                  fitted asymptotic tails.
\end{itemize}

\begin{tabular}{|c|cc|}
\hline
  Type of Calculation &
  ~~$\langle Q_{20}^{proton} \rangle$ [mB]~~~~ &
  $\langle Q_{20}^{neutron} \rangle$ [mB] ~~\\[2mm] \hline
  $^{11}$Li unconstrained HFB & -31.20 & -11.08 \\
  Correlated GCM $^{11}$Li & -31.13 & 2.87   \\
  $^{9}$Li core projected from & & \\ [-5mm]
  $^{11}$Li unconstrained HFB  & -31.20 & -54.94 \\
  $^{9}$Li core projected from & & \\ [-5mm]
  correlated GCM $^{11}$Li  & -31.13 & -53.76 \\
  Free $^{9}$Li, HFB, {\it D1S}  &  -43.46 & -96.24 \\
  $^{11}$Li, experiment \cite{expq20} & -31.2 (4.5) & \\
  $^{9}$Li, experiment \cite{expq20} & -27.4 (1.0) & \\ [2mm] \hline
\end{tabular}

\begin{itemize}
\item[Table III:]  The calculated axial quadrupole moments in millibarns
            for the neutron, proton and total
            matter distributions of $^{11}$Li, $^{9}$Li and the $^{9}$Li
            core projected from $^{11}$Li calculation are listed.  The results
            from the axial HFB and GCM calculations are used.
\end{itemize}

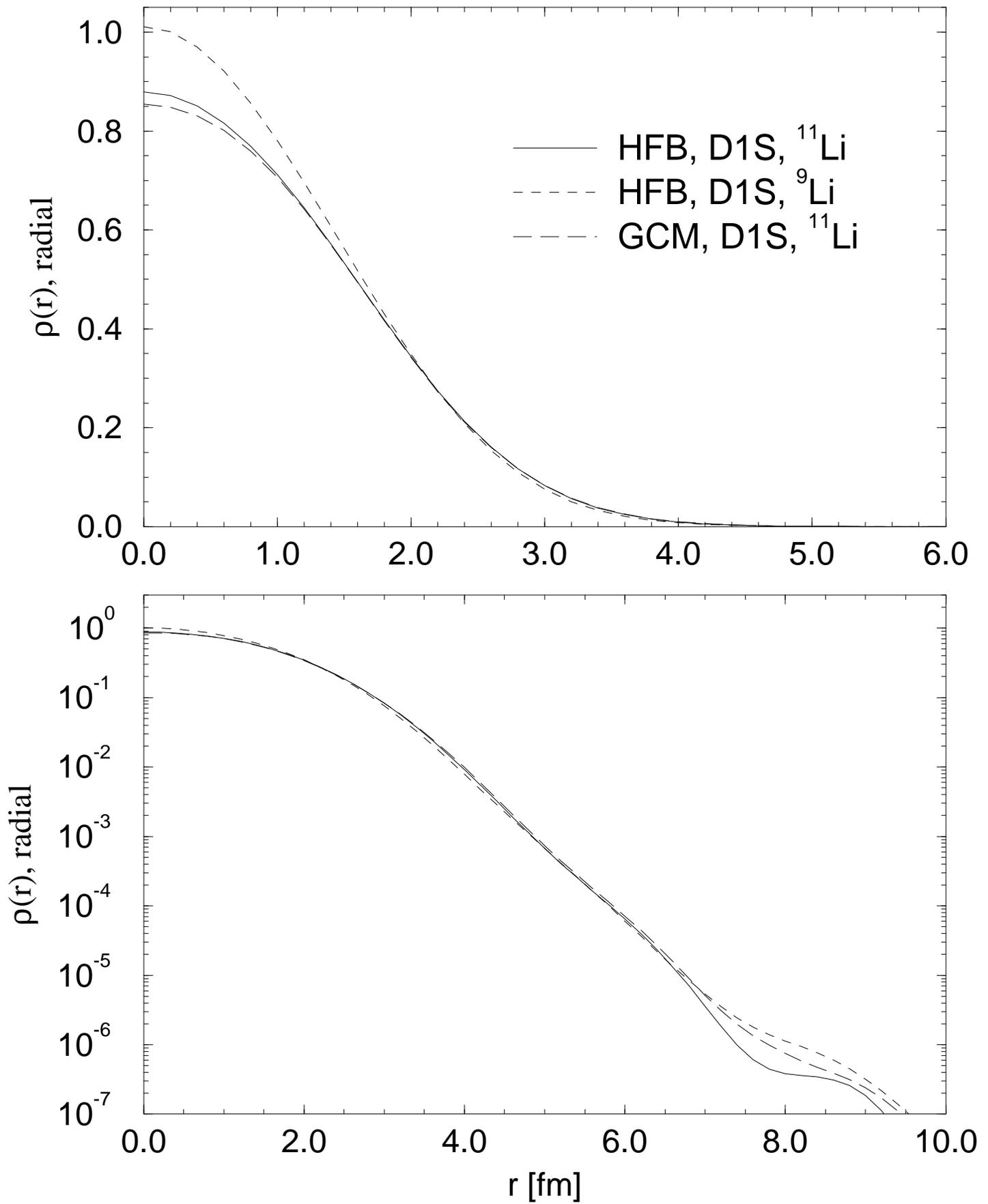
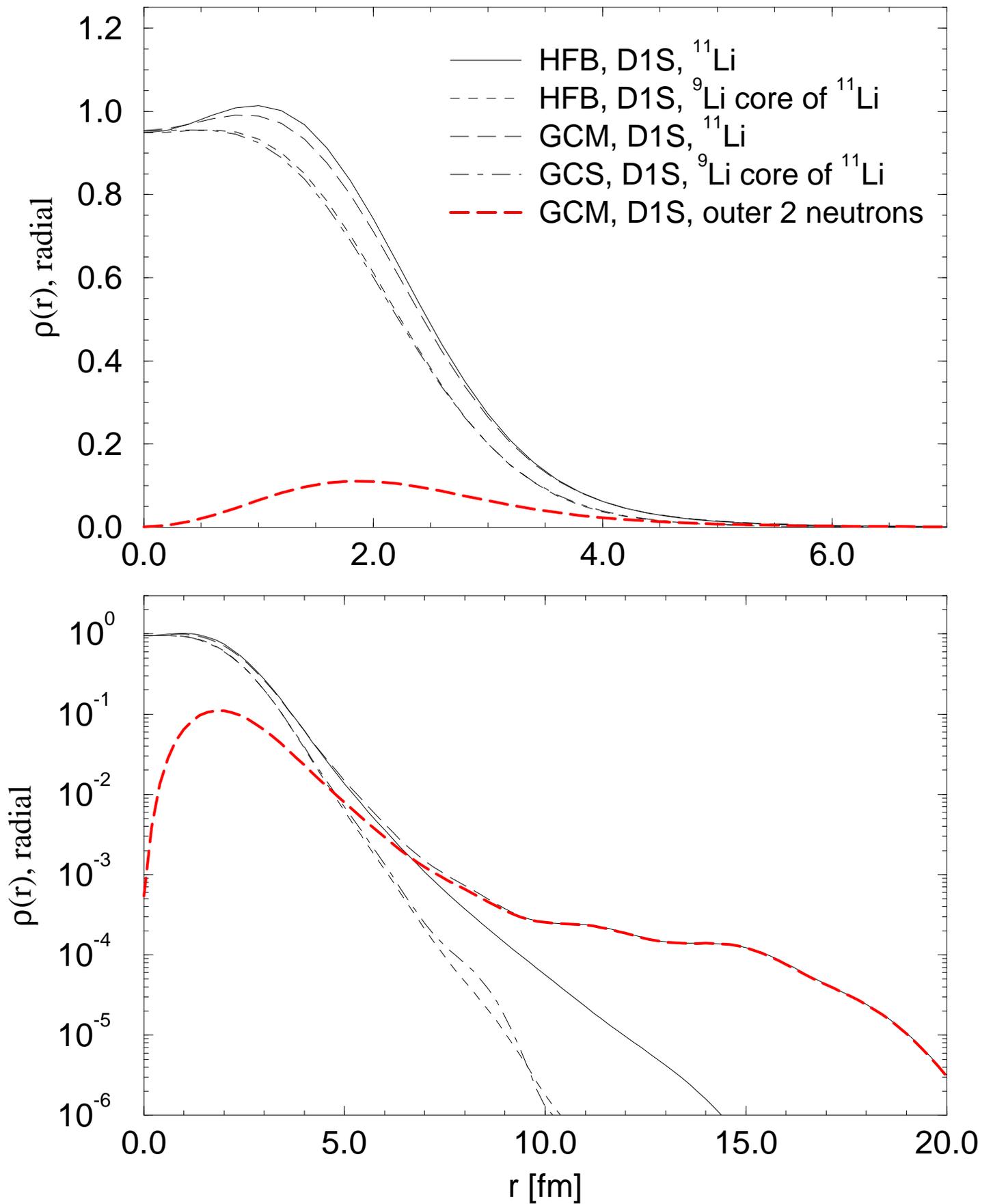
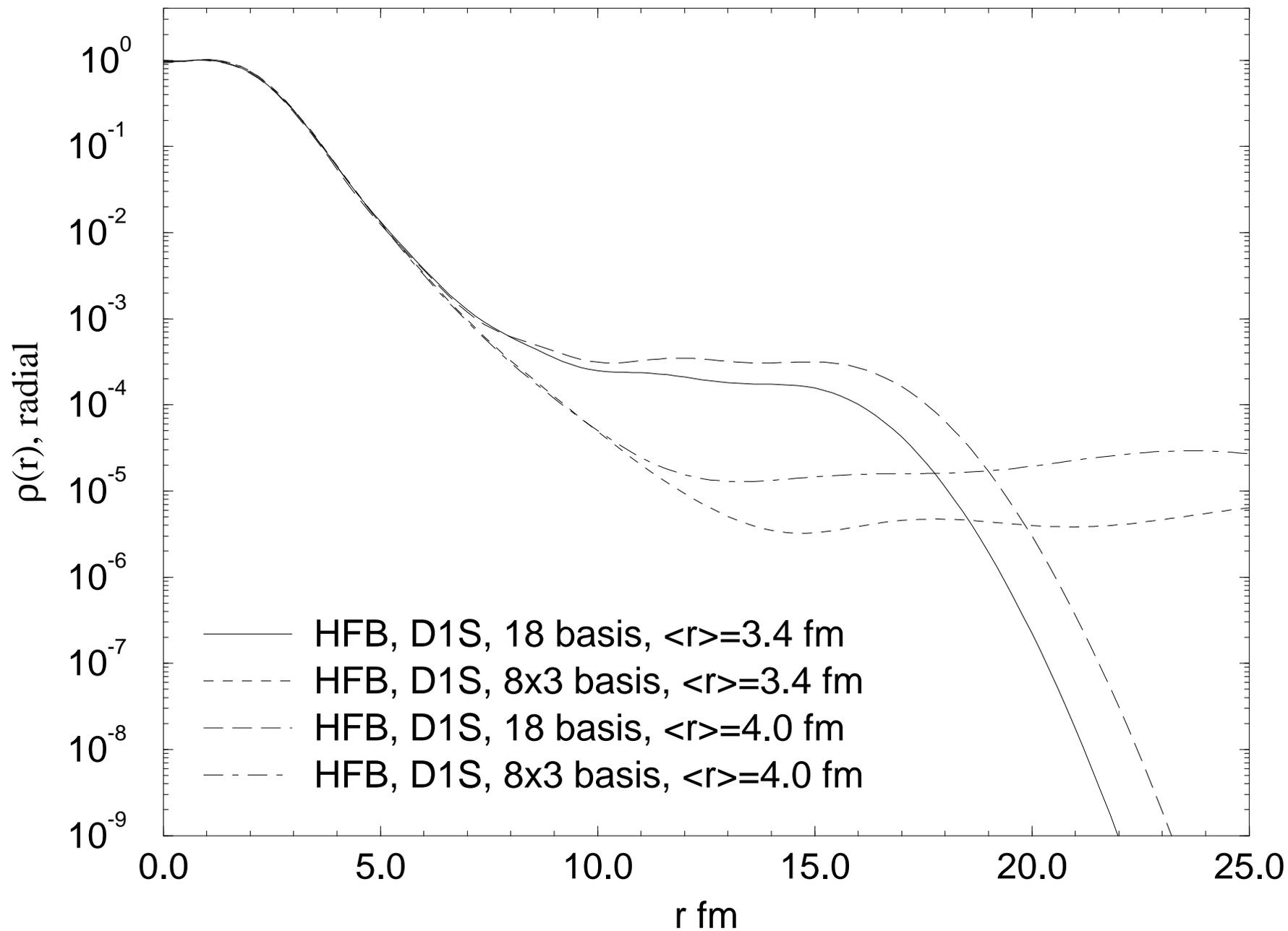
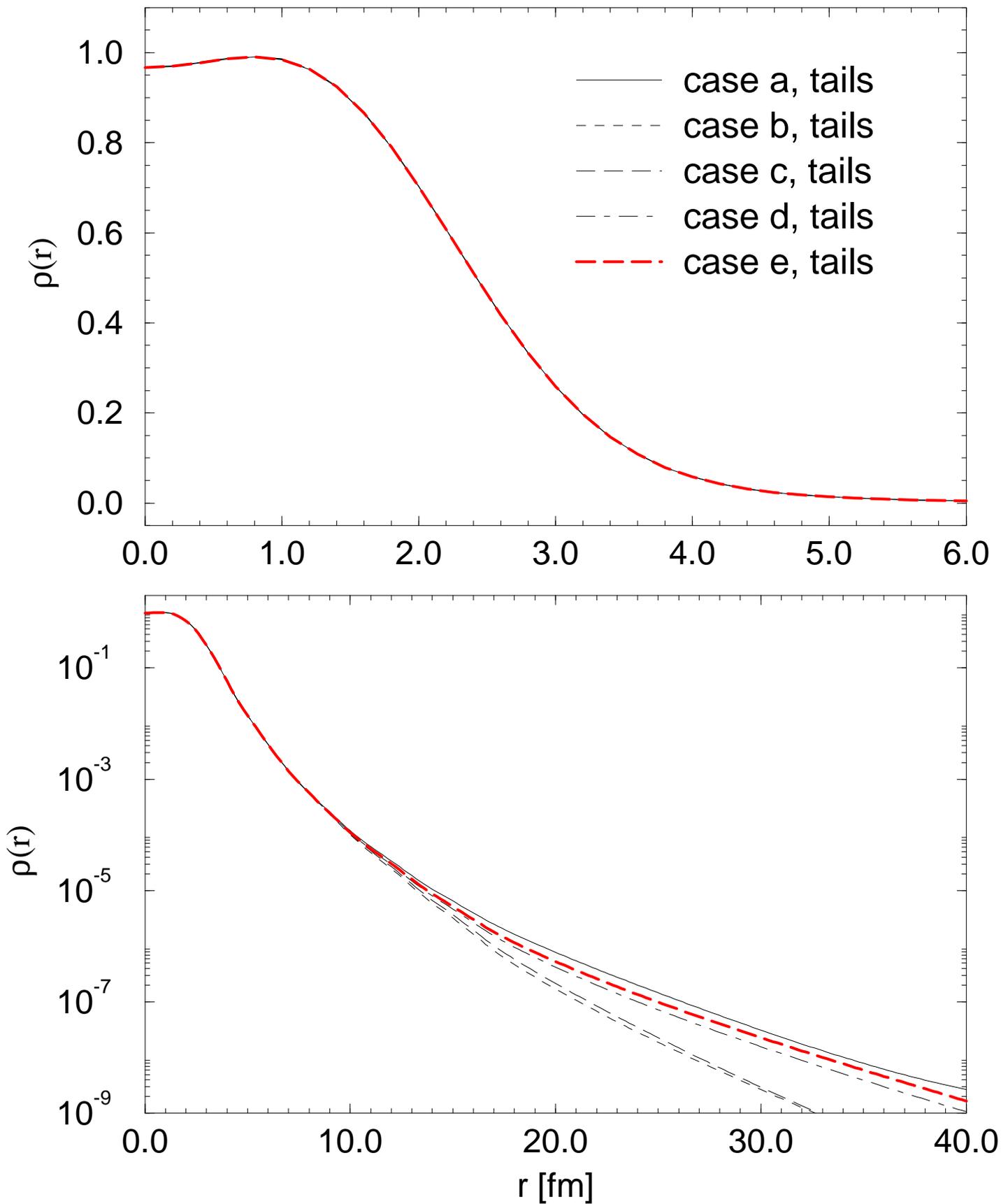
\end{document}